\documentclass[11pt]{article}

\usepackage{graphicx}
\usepackage{color}

\begin{document}

\title{A narrow-band speckle-free light source via random Raman lasing}

\author{Brett H. Hokr$^{a,b \ast}$\thanks{$^\ast$Corresponding author. Email: brett.hokr@tamu.edu\vspace{6pt}},
Morgan S. Schmidt$^c$,
Joel N. Bixler$^c$,\\
Phillip N. Dyer$^b$,
Gary D. Noojin$^b$, 
Brandon Redding$^d$,\\
Robert J. Thomas$^c$, 
Benjamin A. Rockwell$^c$,
Hui Cao$^d$,\\
Vladislav V. Yakovlev$^a$
and Marlan O. Scully$^{a,e,f}$\vspace{6pt}  \\
$^{a}${\em{Texas A\&M University, College Station, TX}};\\
$^{b}${\em{TASC Inc., San Antonio, TX}};\\
$^{c}${\em{711th Human Performance Wing, Human Effectiveness Directorate,}}\\ 
{\em{Bioeffects Division, Optical Radiation Bioeffects Branch,}}\\
{\em{JBSA Fort Sam Houston, TX}};\\
$^{d}${\em{Yale University, New Haven, CT}};\\
$^{e}${\em{Princeton University, Princeton NJ}};\\
$^{f}${\em{Baylor University, Waco, TX}};}

\maketitle

\begin{abstract}
Currently, no light source exists which is both narrow-band and speckle-free with sufficient brightness for full-field imaging applications. Light emitting diodes (LEDs) are excellent spatially incoherent sources, but are tens of nanometers broad. Lasers on the other hand can produce very narrow-band light, but suffer from high spatial coherence which leads to speckle patterns which distort the image. Here we propose the use of random Raman laser emission as a new kind of light source capable of providing short-pulsed narrow-band speckle-free illumination for imaging applications.
\end{abstract}

\section{Introduction}
One of the biggest limitations of optical microscopy techniques for biological applications is speed. Laser based microscopy techniques including  Raman~\cite{Freudiger2008,Yakovlev2010,Beier2011,Ozeki2012,Meng2013,Scully2002,Evans2005,Petrov2007,Arora2012,CampJr2014}, and fluorescence~\cite{Denk1990,Hell1994,Helmchen2005,Horton2013} are limited to scanning pixel by pixel due to the speckle pattern of the laser caused by the large spatial coherence. To acquire high-resolution, megapixel-scale images at video rate (30 frames per second) one needs to have a pixel acquisition time of less than 33~ns. While laser sources with repetition rates in excess of this are readily available often times the signal level, detection equipment, and scanning rates of the laser cannot achieve sufficient speeds. Thus, to obtain real-time, dynamic information about the sample using a laser-based microscopy technique one must sacrifice resolution, signal to noise, field of view, or all of the above.

Traditionally, the way around the issue of speckle is to use an incoherent light source such as an arc lamp~\cite{Lichtman2005} or light emitting diodes (LEDs)~\cite{Albeanu2008} to do full-frame microscopy. However, such sources lack sufficient spectral radiance for Raman spectroscopy, and lack the temporal peak power to be used for any nonlinear optical effects. More recently, random lasing emission and highly multi-mode chaotic cavity lasers have been shown to provide light that is both bright and speckle free~\cite{Redding2012}, but they are still broadband (typically 10's of nanometers). That is, there is no existing light source capable of producing narrowband, nanosecond-duration, speckle-free light. Random lasing via a narrowband Raman transition, known as random Raman lasing (RRL)~\cite{Hokr2014,Hokr2014a,Hokr2014c,Hokr2014d}, provides a bright emission which is narrowband and has low spatial coherence.

Random lasing can be simply thought of as a laser in which feedback is provided by elastic scattering from a powder instead of the mirrors of a Fabry-P\'erot~\cite{Cao1999}. Similar to traditional lasers, random lasers also have modes; however, random lasing modes are not as simple as transverse and longitudinal modes, and are usually more coupled than they would in a typical laser cavity~\cite{Tuereci2008}. This leads to strong mode competition and highly multimode emission. Each of these modes would produce its own independent speckle pattern, similar to passing a single-mode laser through a diffuser or multimode fiber, but because there are many independent modes there are many speckle patterns superimposed which average out to provide a speckle-free emission very well suited for imaging~\cite{Redding2012,Mermillod-Blondin2013}. This highly multimode nature that gives rise to the speckle-free emission also results in most of spontaneous emission bandwidth of the gain medium being used, making random lasers quite broadband. The use of barium sulfate (BaSO$_4$) as the Raman gain medium provides a narrow Raman linewidth of 8~cm$^{-1}$ (0.25~nm at 562~nm), making the RRL emission both narrowband and speckle-free~\cite{Hokr2014,Hokr2014a}. In addition, random Raman lasing can be achieved at a wide range of wavelengths simply by changing the wavelength of the pump laser. 

In this paper, we show that the light emitted from random Raman lasing is an entirely new kind of light source capable of producing very bright, narrowband, temporally fast, and most importantly speckle-free light. Such a source opens the possibility of a low-spatial-coherence light source while simultaneously allowing a relatively long temporal coherence due to the narrow bandwidth. Additionally, we will demonstrate stroboscopic images of microcavitation of melanasomes.

\section{Results and Discussion}

\begin{figure}[tb]
\centering
\includegraphics[width=0.6\textwidth]{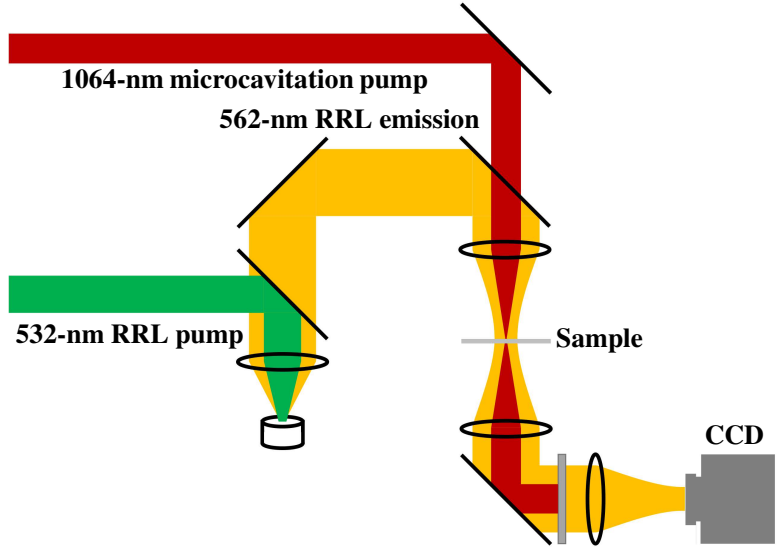}
\caption{Schematic diagram of the experimental setup.}
\label{fig:Schematic}
\end{figure}

In order to verify that RRL is indeed speckle-free and may be used for strobe microscopy, the formation of bovine melanosome microcavitation bubbles were imaged.  Melanin-containing organelles known as melanosomes are the main absorbing particles found in the monolayer of cells of the retinal pigment epithelium (RPE) layer of the retina~\cite{Lin1998,Lee2007}. The formation of small bubbles (microcavitation) is the primary retinal damage mechanism for laser pulses in the $10^{-9}$ to $10^{-6}$ seconds range around the melanosomes in the RPE cells~\cite{Lund2014,Brinkmann2000}. The formation of cavitation bubbles are highly correlated to cell death, and it is critical to understand the relationship between laser irradiance and damage. By using strobe microscopy, one can determine damage thresholds by imaging the cavitation events. Probit analysis, the standard technique for ascribing threshold values, was used to estimate the laser-induced damage (ED50) required to cause damage through microcavitation images~\cite{Cain1996,Finney1971}. Additional experimental details pertaining to the traditional use of strobe photography in microcavitation studies are discussed in Schmidt et al.~\cite{Schmidt2014}.

\begin{figure}[tb]
\centering
\includegraphics[width=0.8\textwidth]{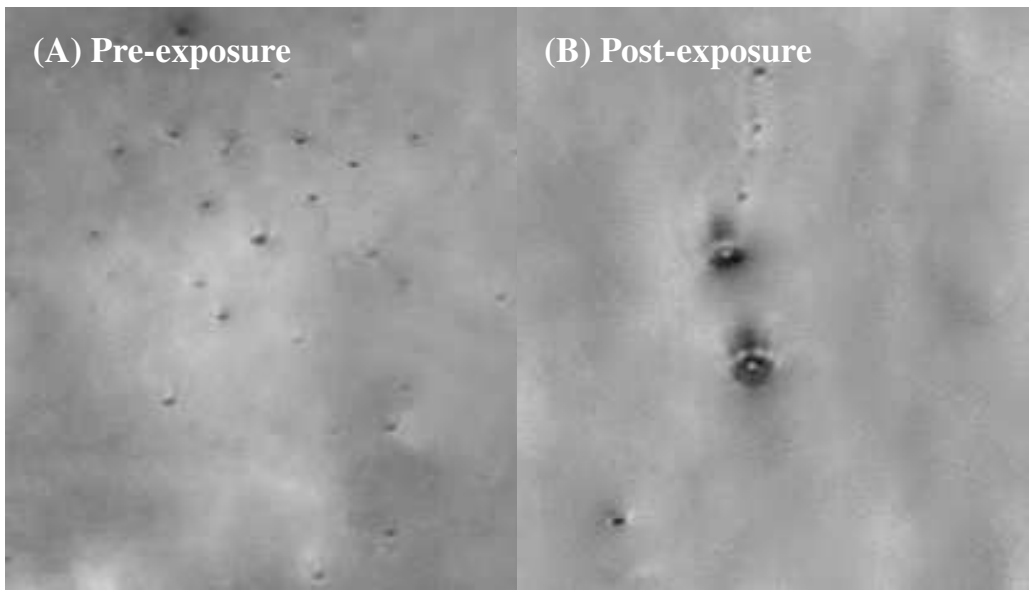}
\caption{(A) Isolated bovine melanasomes before irradiation with the 1064-nm pulse. (B) Background subtracted melanasomes 125~ns after irradiation displaying the formation of microcavitation bubbles. Both of these images were illuminated using random Raman lasing emission.}
\label{fig:Microcavitation}
\end{figure}

For this experiment, shown schematically in Fig~\ref{fig:Schematic}, two laser systems were needed in order to observe the microcavitation events after exposure. First, the RRL, made of BaSO$_4$ powder with particle diameters of 1-5~$\mu$m, was pumped with a frequency-doubled Spectra Physics Model GCR-130 Nd:YAG for an output of 532~nm at 10~Hz. The 532~nm beam was gently focused on the BaSO4, with an approximate spot size of 1~mm. This optical spot size was determined through a previous study~\cite{Hokr2014}. A lens was used to collimate the 562~nm emission from the BaSO$_4$ at a 45$^{\circ}$ angle.  The beam was then directed towards the melanosome sample through a set of mirrors and through the back of an infrared mirror that turned the irradiation beam to the sample. Second, the illumination beam was co-aligned with the irradiation beam, which consisted of a 10~ns pulse duration from a 1064~nm Nd:YAG (spectra physics, INDI-30). Time-resolved imagery was achieved by using a delay generator to control the delay between the irradiation beam and the illumination beam, as well as trigger the Bobcat CCD camera (Imperx Incorporated, Boca Raton, Florida) to capture an image.  A long-working-distance microscope 10x objective was used with a 400-mm tube lens for a total magnification of 20x~\cite{Schmidt2014},providing micrometer spatial resolution and a nanosecond strobe with steps of nanoseconds relative to the laser illumination. Furthermore, the experimental setup allowed for the observation of microcavitation formation, as well as pre- and post-cavitation images.  Background subtraction from the exposure image allowed for improved photographs of the cavitation events. The results are shown in Fig.~\ref{fig:Microcavitation}.

To more fully explore the spatial coherence properties of RRL emission we have preformed a Young's double-slit experiment illuminated with three different light sources, RRL emission, elastically scattered 532~nm radiation from the pump laser, and a Helium-Neon laser for reference. In this experiment the RRL was pumped with a 50-ps pulse from a Spectra Physics Quanta-Ray GCR-3RA. The pulses had a pulse energy of 530~$\mu$J and were gently focused to a beam diameter of 0.83~mm on the surface of the powder corresponding to an intensity approximately three times larger than the threshold required for random Raman lasing. The surface of the BaSO$_4$ powder was imaged to the double-slit with 10x magnification in a 4-f arrangement. The two slits were 100~$\mu$m wide and separated by 350~$\mu$m. Following the slit a cylindrical lens was used to image the vertical direction onto the CCD with unit magnification. Thus, this setup effectively probes the spatial coherence of two 10~$\mu$m thick lines on the surface of the powder separated by 35~$\mu$m. For the elastically scattered pump the laser was attenuated to well below the lasing threshold to avoid artifacts stemming from pump depletion. The Helium-Neon laser was obtained with the same lens arrangement to allow direct comparison.

\begin{figure}[tb]
\centering
\includegraphics[width=0.95\textwidth]{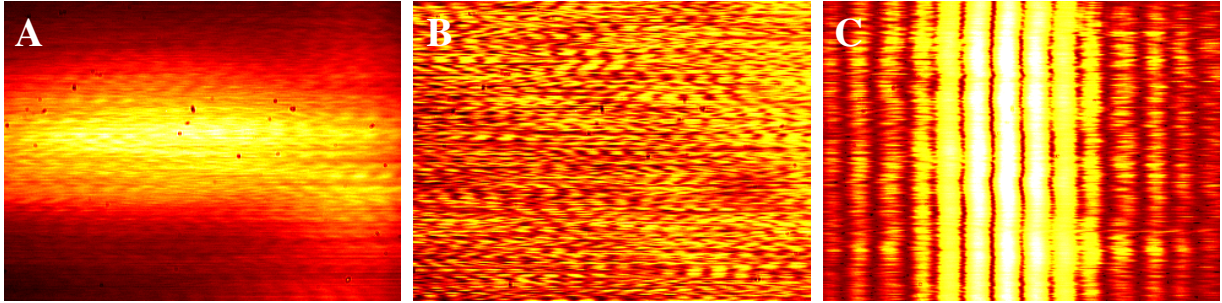}
\caption{Double-slit diffraction pattern generated by (A) random Raman laser emission, (B) elastically scattered 532~nm pump, and (C) Helium-Neon laser.}
\label{fig:DoubleSlit}
\end{figure}

The results of the double-slit experiment are shown in Fig.~\ref{fig:DoubleSlit}. They clearly show that while a small amount of spatial coherence persist in the RRL emission it is much less pronounced than even that of the elastically scattered pump radiation. This result can be understood most easily by thinking about it in terms of speckle patterns. If a single-spatial-mode laser is passed through a diffuser it will generate a speckle pattern, but if this diffuser is rotated these speckle patters will average out over time to provide a uniform illumination. Each one of these speckle patterns can be thought of as a single mode of the laser diffuser system. In a highly-multimode random laser, including the random Raman laser, each spatial mode of the laser generates a different speckle pattern for a given diffuser (elastic scattering in a random laser). These many speckle patterns average over all to provide a low-coherence source.

\begin{figure}[tb]
\centering
\includegraphics[width=0.9\textwidth]{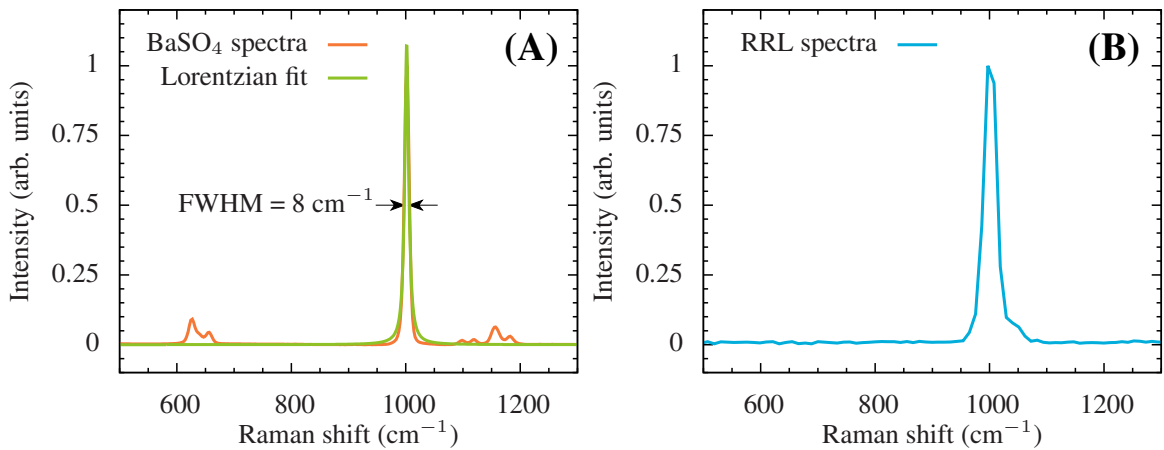}
\caption{(A) Linewidth of the spontaneous Raman spectrum for BaSO$_4$. The Lorentzian fit give the full-width at half-maximum width to be 8~cm$^{-1}$ (0.25~nm at 562~nm). (B) Spectrum of RRL emission. Note that this data set was taken with a lower-resolution spectrometer such that the width of this peak is determined by the resolution of that spectrometer and not by the emission.}
\label{fig:Linewidth}
\end{figure}

In addition to strobe photography applications, RRL emission could have applications as an spatially-incoherent source for spectroscopy. Figure~\ref{fig:Linewidth} shows the spontaneous Raman spectrum for BaSO$_4$ has a spontaneous emission bandwidth of 8~cm$^{-1}$ for the strongest Raman line. This linewidth convoluted with the pump laser spectrum (in this case it is sufficiently narrow that it can be neglected) provides the maximum possible bandwidth of the RRL much like the spontaneous emission bandwidth would determine the maximum gain bandwidth of a laser. Additionally, the RRL emission spectrum is shown to illustrate the lack of the weaker Raman peaks seen in the spontaneous spectrum. The wider width of the RRL emission spectrum here is due to the fact that a lower resolution spectrometer was used to acquire this data. Thus the width of this feature is determined solely by the resolution of the spectrometer and not by the RRL emission. The narrow RRL linewidth would be sufficient for full-field Raman spectroscopy. Given the speckle-free nature combined with the narrow line-width of the RRL emission it would in principle be possible to acquire an entire Raman spectral image in a single laser shot. Even if many shots are required to obtain the required signal to noise this would likely still far exceed the speed at which current imaging Raman microscopes can obtain spectroscopic images. The obvious limitation for such a technique is the lack of a spectrometer capable of detecting a Raman spectrum from each point in a 2-dimensional array. Currently, no such detector exists; however, recent advances in compressive sensing offer hope in this area~\cite{Gao2014}. 

\section{Conclusion}
We have shown that random Raman laser emission is a unique source of light. It can be made very bright (a few percent of the pump energy), it is sufficiently narrowband for spectroscopic applications with a linewidth of at most 8~cm$^{-1}$, and, most importantly, it is sufficiently spatially incoherent such that speckle free full-frame images can be obtained. Additionally, we have acquired proof of principle images demonstrating speckle-free strobe photography images of microcavitation in bovine melanasomes using random Raman laser emission as the stroboscopic illumination source.

\section*{Ackowledgements}
This work was partially supported by National Science Foundation Grants ECCS-1250360, DBI-1250361, CBET-1250363, PHY-1241032 (INSPIRE CREATIV), PHY-1068554, EEC-0540832 (MIRTHE ERC), DMR-1205307 and the Robert A. Welch Foundation (Award A-1261). BHH would like to acknowledge a graduate fellowship from the Department of Defense Science, Mathematics and Research for Transformation (SMART) fellowship program. JNB and RJT would like to acknowledge support from AFRL 711 HPW. Research performed by TASC Inc. and Nanohmics Inc. was conducted under USAF Contract Number FA8650-14-D-6519.

\bibliographystyle{tMOP}
\bibliography{../../../../../library}

\end{document}